\documentclass[12pt]{article}
\usepackage{cite}
\newcommand{\CR}{{\mathcal R}}

\newcommand{\ba}{\begin{array}}
\newcommand{\ea}{\end{array}}
\newcommand{\be}{\begin{equation}}
\newcommand{\ee}{\end{equation}}
\newcommand{\bea}{\begin{eqnarray}}
\newcommand{\eea}{\end{eqnarray}}

\newcommand{\p}{\partial}

\def\CL{{\mathcal{L}}}

\def\CN{{\mathcal{N}}}
\def\IB{\relax\hbox{$\inbar\kern-.3em{\rm B}$}}
\def\IC{\relax\hbox{$\inbar\kern-.3em{\rm C}$}}
\def\ID{\relax\hbox{$\inbar\kern-.3em{\rm D}$}}
\def\IE{\relax\hbox{$\inbar\kern-.3em{\rm E}$}}
\def\IF{\relax\hbox{$\inbar\kern-.3em{\rm F}$}}
\def\IG{\relax\hbox{$\inbar\kern-.3em{\rm G}$}}
\def\IGa{\relax\hbox{${\rm I}\kern-.18em\Gamma$}}
\def\IH{\relax{\rm I\kern-.18em H}}
\def\IK{\relax{\rm I\kern-.18em K}}
\def\IL{\relax{\rm I\kern-.18em L}}
\def\IP{\relax{\rm I\kern-.18em P}}
\def\IR{\relax{\rm I\kern-.18em R}}
\def\IZ{\relax{\rm Z\kern-.5em Z}}

\def\half{\frac{1}{2}}
\def\p{\partial}
\def\f{\frac}

\begin{document}

\begin{titlepage}

\begin{flushright}
hep-th/0302075 \\
February 2003

\end{flushright}

\vskip 2 cm

\begin{center}
{\LARGE Extended chiral algebras and the emergence of $SU(2)$ quantum numbers in the Coulomb gas} \vskip
1 cm { \large A. Nichols\footnote{nichols@inp.demokritos.gr} }

\begin{center}
{\em National Research Centre Demokritos, \\
Institute of Nuclear Physics, \\
Agia Paraskevi, \\
GR-15310 Athens, Greece.}
\end{center}

\vskip 1 cm

\vskip .5 cm

\begin{abstract}
We study a set of chiral symmetries contained in degenerate operators beyond the `minimal' sector of the $c_{p,q}$ models. For the operators $h_{(2j+2)q-1,1}=h_{1,(2j+2)p-1}$ at conformal weight $\left[ (j+1)p-1 \right] \left[ (j+1)q -1 \right]$, for every $2j \in \mathbf{N}$, we find $2j+1$ chiral operators which have quantum numbers of a spin $j$ representation of $SU(2)$. We give a free-field construction of these operators which makes this structure explicit and allows their OPEs to be calculated directly without any use of screening charges. The first non-trivial chiral field in this series, at $j=\half$, is a fermionic or parafermionic doublet. The three chiral bosonic fields, at $j=1$, generate a closed $W$-algebra and we calculate the vacuum character of these triplet models.
\end{abstract}

\end{center}

\end{titlepage}

\section{Introduction}
Conformal field theory (CFT) has attracted immense attention in both the mathematics and physics literature since the seminal work of Belavin, Polyakov, and Zamolodchikov \cite{Belavin:1984vu}. In CFT a crucial role is played by the chiral
algebra of the theory and all fields transform in representations
of this algebra. A special class of theories called rational CFTs
is of particular importance as they contain only a finite number
of basic representations. In normal CFT these representations are
all irreducible. 

The most common chiral algebra is the Virasoro algebra. The rational
theories that can be built using this algebra alone are known as
the Virasoro minimal models. They are characterised by a discrete
set of central charges $c_{p,q}$ (we will always consider models with $p,q \in \mathbf{N}$ and take $gcd(p,q)=1$):
\bea 
\label{eqn:Vircentralc}
c_{p,q}=1-6 \f{(p-q)^2}{pq} \eea
and the following set of representations:
\bea
\label{eqn:Kactableweights}
h_{r,s}=\f{(p r-q s)^2-(p-q)^2}{4pq}
\eea
with $1 \le r \le q-1$ and $1 \le s \le p-1$ and the identifications
$h_{r,s}=h_{q-r,p-s}$. The operators within this set form a closed
algebra under fusion giving us a rational CFT. The correlators of the degenerate fields in these models can be found by solving differential equations or by the use of the Coulomb gas representation \cite{Dotsenko:1984nm}.
\subsection{Beyond the minimal sector}
One may consider the question of what happens if one considers including some
degenerate fields from beyond this minimal sector. This is not purely an academic issue as operators beyond the minimal trivial sector of $c_{2,3}=0$ are known to be connected with percolation \cite{Cardy:1992cm}. These operators
still possess null vectors allowing us to calculate the conformal
blocks. In contrast to the minimal sector, we will no
longer have closure of the operator algebra and continued fusion 
will generate an infinite set of operators leading us out of the space of rational
CFTs. However if there is a larger chiral algebra present then there may again exist the possibility of re-arranging the operators into a finite set. Finding such chiral algebras will be the subject of this paper.

Many of the correlators of irreducible operators beyond the minimal sector have logarithmic singularities. In such theories, known as Logarithmic Conformal Field theories (LCFTs), the appearance of logarithms is a signal that the irreducible primary operators do not close under fusion and indecomposable representations are inevitably generated \cite{Gurarie:1993xq}. The appearance of logarithms was first observed in the context of WZNW models based on supergroups \cite{Rozansky:1993td,Rozansky:1992rx}. The Coulomb gas approach can in principle be used to calculate correlators in these theories. However in practice it requires a very careful combination of singular integrals and it is much more convincing to solve the necessary differential equations directly. 

A great deal of work has been done on analysing LCFTs and their applications in many different contexts for example: WZNW models and gravitational dressing \cite{Bilal:1994nx,Caux:1997kq,Kogan:1997nd,Kogan:1997cm,Giribet:2001qq,Gaberdiel:2001ny,Nichols:2001du,Kogan:2001nj,Nichols:2001cv,Lesage:2002ch},
polymers \cite{Saleur:1992hk,Cardy,Gurarie:1999yx}, disordered systems and the
Quantum Hall effect
\cite{Caux:1996nm,Kogan:1996wk,Maassarani:1997jn,Gurarie:1997dw,CTT,Caux:1998eu,Bhaseen:1999nm,Kogan:1999hz,Gurarie:1999bp,RezaRahimiTabar:2000qr,Bernard:2000vc,Bhaseen:2000bm,Bhaseen:2000mi,Ludwig:2000em,CardyTalk,Kogan:2001ku},
string theory
\cite{Kogan:1996df,Kogan:1996zv,Ellis:1998bv,Ghezelbash:1998rj,Kogan:1999bn,Myung:1999nd,Lewis:1999qv,Nichols:2000mk,Kogan:2000nw,Moghimi-Araghi:2001fg,Bakas:2002qh,Jabbari-Faruji:2002xz},
2d turbulence
\cite{RahimiTabar:1996dh,RahimiTabar:1997nc,Flohr:1996ik,RahimiTabar:1997ki,RahimiTabar:1996si},
multi-colour QCD at low-x \cite{Korchemsky:2001nx}, the Abelian sandpile
model\cite{Mahieu:2001iv,Ruelle:2002jy} and the Seiberg-Witten solution of
${\mathcal N}=2$ SUSY Yang-Mills \cite{Cappelli:1997qf,Flohr:1998ew}. Deformed
LCFTs, Renormalisation group flows and the $c$-theorem were discussed in
\cite{Caux:1996nm,Rahimi-Tabar:1998ph,Mavromatos:1998sa}. The holographic relation
between logarithmic operators and vacuum instability was considered in
\cite{Kogan:1998xm,Lewis:1998fg}. There has also been much interest on LCFTs with a
boundary \cite{Kogan:2000fa,Ishimoto:2001jv,Kawai:2001ur,Bredthauer:2002ct}. For
more about the general structure of LCFT see
\cite{RahimiTabar:1997ub,Rohsiepe:1996qj,Kogan:1997fd,Flohr:1998wm,Flohr:2000mc,Flohr:2001tj} and references
therein. Introductory lecture notes on LCFT and more references can be found in
\cite{Tabar:2001et,Flohr:2001zs,Gaberdiel:2001tr,Moghimi-Araghi:2002gk}. A general
approach to LCFT via deformations of the operators has been suggested in \cite{Fjelstad:2002ei} - see also \cite{Mavromatos:2002fm,Krohn:2002gh}.

Our attention here will be on the $c_{p,q}$ theories outside the minimal sector although we expect that much of what we say may hold in more generality. The $c_{p,1}$ models, and in particular the case of $c_{2,1}=-2$, have received considerable attention in the literature due to the discovery that one may extend the Virasoro algebra by triplets of chiral $h_{3,1}=2p-1$ fields \cite{Kausch:1991vg}. The resulting algebra is sufficient is create a rational LCFT, i.e. one having only a finite number of
irreducible and indecomposable representations \cite{Kausch:1995py,Flohr:1996ea,Gaberdiel:1996np,Flohr:1997vc,Gaberdiel:1998ps,Kausch:2000fu}. Beyond the $c_{p,1}$ series little is known and we hope that the results here may help to shed some light on the issue.
                                
As we wish to consider theories based on the operators in the
Kac-table it is natural that the additional algebras
should themselves be generated by a subset of the degenerate
operators. Such operators must satisfy the condition
that they can consistently be considered chiral
within correlation functions.

In particular let us consider the correlation function of four such operators that are degenerate at the same level:
\bea \label{eqn:fourpointfn} 
\hspace{-0.3cm}\langle \phi(z_1)
\phi(z_2) \phi(z_3) \phi(z_4) \rangle =z_{42}^{-2h}
 z_{31}^{-2h}  F(z)
\eea
where the $F(z)$ is constructed from the conformal blocks $F^{(i)}(z)$ and the cross ratio $z$ is given by:
\bea
z=\frac{z_{12}z_{34}}{z_{13}z_{24}} 
\eea
A necessary condition for the existence of a chiral algebra is the
appearance of a rational function, that is a function with only a finite number of poles, within the conformal blocks
\footnote{The appearance of a rational solution is so restrictive
that we do not know of examples where it is not sufficient.}. The fact that a rational
function can be reconstructed from its pole structure and behaviour at infinity is in exact coincidence with the fact that chiral algebras are determined by
the singular terms in the OPE. Note that as we wish to consider possible chiral operators we do not have to combine the holomorphic and anti-holomorphic conformal blocks in (\ref{eqn:fourpointfn}). Here we shall discuss a particularly interesting series of chiral algebras present in the $h_{1,s}$ fields. The $h_{1,s}$ fields beyond the minimal sector are certainly good candidates for extensions of the chiral algebra as they close amongst themselves under fusion (by which we shall always mean on the sphere). It would be interesting to understand more completely the appearance of chiral algebras generated by fields within the Kac-table. In \cite{Nichols:2002dk} by explicit consideration of a large number of examples we found that the $h_{1,(2j+2)p-1}$ and $h_{(2j+2)q-1,1}$ operators\footnote{If we naively attempted to consider a $c_{(2j+2)p,(2j+2)q}$ model, with $1 \le r \le (2j+2)q-1, 1\le s \le (2j+2)p-1$, then these operators would lie at two of the corners. However it is not clear if there is any precise meaning to such a description and therefore we shall not make use of it in the following.} (both having the same dimension $\left[ (j+1)p-1 \right] \left[ (j+1)q -1 \right]$) always possessed an identical subset of $2j+1$ rational correlation functions. The fact that a
subset of their conformal blocks agree is already a non-trivial
statement as these fields possess null vectors at different
levels. Based on the behaviour under crossing symmetry of these rational solutions we conjectured in \cite{Nichols:2002dk} that the $2j+1$ degeneracy was due to a multiplet of chiral fields having \emph{extra} $SU(2)$ quantum numbers.

As an example consider the $h_{1,3}=1$ operators from the $c_{1,1}=1$ model. Solving a third order differential equation for (\ref{eqn:fourpointfn}) we find the following conformal blocks:
\bea 
F^{(1)}&=& \f{z^2}{(1-z)^2} \nonumber \\
F^{(2)}&=& \f{(1-z+z^2)^2}{z^2(1-z)^2} \\
F^{(3)}&=&\f{(1-z)^2}{z^2} \nonumber
\eea
The appearance of these three 
rational conformal blocks can be explained by the fact that at 
$h=1$ there is not one but \emph{three} chiral primary fields which we shall call $W^a(z)$. These fields are \emph{identical} as far as the Virasoro algebra is concerned as they have null vectors at the same level. In order to distinguish them we introduce an extra quantum number. This is not arbitrary and must be consistent with the transformation of the conformal blocks under crossing symmetry. The correct choice is:
\bea 
\label{eqn:c11triplet}
\left< W^+(0) W^+(z) W^-(1) W^-(\infty) \right>&=& F^{(1)} \nonumber\\
\left< W^3(0) W^3(z) W^3(1) W^3(\infty) \right>&=&  F^{(2)} \\
\left< W^+(0) W^-(z) W^-(1) W^+(\infty) \right>&=& F^{(3)}\nonumber 
\eea
In this example these chiral algebra fields are well known and are simply the currents $J^a(z)$ of $SU(2)_1$. The $SU(2)_1$ model is certainly not an LCFT (at least if we are restricted within the integrable representations) however it serves to illustrate the general arguments. In more complicated examples we can analyse the behaviour of the rational conformal blocks under crossing symmetry to \emph{deduce} the possible multiplet nature of the chiral fields. In this paper we will provide a free field construction for this subset of chiral primary operators, which we shall call $\Phi_{(j)}$, with conformal weight:
\bea
h_{(j)}=\left[(j+1)p-1 \right] \left[(j+1)q-1 \right]
\eea
We shall explicitly demonstrate their chiral nature and extra $SU(2)$ multiplet structure\footnote{Here we restrict ourselves with $p,q \in \mathbf{N}$ as although a similar series of rational solutions can be found in the $c_{-p,q}$ cases the degenerate primary fields are of increasingly negative dimension and it is not clear how one should proceed.}. However before we do so let us describe the standard Coulomb gas approach \cite{Dotsenko:1984nm}.
\section{Coulomb gas approach}
We begin with a free boson normalised in the standard way:
\bea 
\phi(z) \phi(w) \sim -\ln(z-w) 
\eea
We form the stress tensor:
\bea 
T=-\half \p \phi \p \phi + i \sqrt{2} \alpha_0 \p^2 \phi 
\eea
which has the central charge:
\bea 
c=1-24 \alpha_0^2 
\eea
Now with the choice:
\bea \label{eqn:alphazero}
\alpha_0=\f{p-q}{2 \sqrt{p q}} 
\eea
we reproduce the central charge of the $c_{p,q}$ model (\ref{eqn:Vircentralc}). One may introduce vertex operators of the form:
\bea 
V_{\alpha}(z)=e^{i \sqrt{2} \alpha \phi(z)} 
\eea
and these have dimension:
\bea 
h=\alpha^2-2\alpha \alpha_0 
\eea
Non-vanishing correlators must satisfy the charge neutrality condition:
\bea
\label{eqn:chargeneut}
\sum_{k} \alpha_k = 2 \alpha_0
\eea
In order to satisfy this one first introduces the screening charges which are integrals over dimension one operators:
\bea \label{eqn:CGscreenings}
Q_{\pm}= \oint dz~ V_{\pm}  = \oint dz~ e^{i \sqrt{2} \alpha_{\pm} \phi(z)}
\eea
where $\alpha_{\pm}$ satisfies $\alpha_{\pm}^2-2 \alpha_{\pm} \alpha_0 =1$ and with our choice of $\alpha_0$ (\ref{eqn:alphazero}) we have $\alpha_+=\sqrt{p/q},\alpha_-=-\sqrt{q/p}$. As the $Q_{\pm}$ operators commute with the stress tensor they do not affect the conformal properties but they will screen the charge allowing us to satisfy the charge neutrality constraint (\ref{eqn:chargeneut}). There are now a special set, known as admissible charges:
\bea \label{eqn:CGalphas}
\alpha_{r,s}=\half (1-r) \alpha_+ + \half (1-s) \alpha_-=\f{1}{2 \sqrt{pq}} \Bigl\{ p(1-r) -q(1-s) \Bigr\}
\eea
which give rise to operators with singular vectors in their representations. The dimensions of these reproduce the entries of the Kac-table $h_{r,s}$.

However in practice beyond the `minimal' sector such an integral representation is of very little use as one must take a very careful limiting combination of correlators in order to obtain non-trivial results. In addition to this there is a conceptual point - there is \emph{no} space for any extra quantum numbers in this prescription.

We shall now see that to calculate correlators of a subset of the fields, namely the $\Phi_{(j)}$, with conformal weight $\left[(j+1)p-1 \right] \left[ (j+1)q -1 \right]$ there is another prescription in which, with a different representation of the operators, no integrations are necessary and the OPEs can be calculated directly. Moreover the chiral fields come in a $2j+1$ dimensional representation of $SU(2)$.
\subsection{Some numerology}
We shall begin by showing that fields with conformal weight $\left[
(j+1)p-1 \right] \left[ (j+1)q -1 \right]$ can be expressed as
descendants of other Kac-table operators in exactly $2j+1$
different ways. Therefore we try to solve the equation:
\bea 
\label{eqn:main1} 
h_{r,s}+rs=\left[ (j+1)p-1 \right] \left[
(j+1)q -1 \right] 
\eea
with $r,s \in {\mathbf{N}}$. This equation has a very simple set of solutions.
First we solve for $r$:
\bea 
\f{(pr-qs)^2-(p-q)^2}{4pq}+rs &=&
\left[ (j+1)p-1 \right]\left[(j+1)q -1 \right] \nonumber \\
\f{(pr+qs)^2-(p-q)^2}{4pq} &=&
\left[ (j+1)p-1 \right]\left[(j+1)q -1 \right]  \nonumber\\
(pr+qs)^2 &=& 4pq \left[ (j+1)p-1
\right]\left[(j+1)q -1 \right] +(p-q)^2 \\
(pr+qs)^2 &=& \left[ p+q -2(j+1)pq \right]^2  \nonumber\\
 pr &=& -qs \pm \left[p+q -2(j+1)pq \right]  \nonumber\\
 r &=& \f{q}{p} \left( \pm 1 -s \right) \pm\left[1 -2(j+1)q \right]
 \nonumber 
 \eea
It is easy to see that we must take the negative sign as otherwise
$r$ is not positive. We therefore have the solution:
\bea 
r &=& \f{q}{p} \left( - 1 -s \right) - \left[1 -2(j+1)q \right] 
\eea
Now as $gcd(p,q)=1$ we must have $-1-s=p ~{\mathbf{Z}}$. For a
reason that will become clear very shortly we shall rewrite this
as:
\bea 
s=(j+1+m)p-1 
\eea
where $m=-j+ {\mathbf{Z}}$ and $j$ is for now arbitrary. Now substituting the solution for $r$ we get:
\bea 
r&=&\f{q}{p} \left( -1 -(j+1+m)p+1 \right) - \left[ 1-2(j+1)q \right] \\
\Rightarrow r&=&(j+1-m)q-1  \nonumber 
\eea
Therefore the solution set to (\ref{eqn:main1}) is given by:
\bea 
\label{eqn:solns}
r &=& (j+1-m)q-1 \\
s &=& (j+1+m)p-1  \nonumber 
\eea
and for positive integer solutions of both $r$ and $s$ we can only have $2j\in \mathbf{N}$ and $m=-j,-j+1,\cdots j$. In other words there are exactly $2j+1$
solutions to equation (\ref{eqn:main1}). One should take some care in the cases in which $p$ and/or $q$ is equal to one and we shall comment on this below.

We also observe that with the solutions (\ref{eqn:solns}) we have:
\bea 
\label{eqn:hrssolns} 
h_{r,s}&=&\left[ (j+1)p-1 \right] \left[ (j+1)q -1 \right]- \left[(j+1-m)q-1 \right] \left[(j+1+m)p-1 \right] \nonumber \\
&=& m^2pq+m(p-q) 
\eea
Now we see that the choice $\alpha=-m \sqrt{p~q}$ we get a vertex operator with precisely the dimensions of the operators $h_{r,s}$ in
(\ref{eqn:hrssolns}).

Therefore our fields $\Phi_{(j)}$ at conformal weight $h_{(j)}=\left[(j+1)p-1 \right] \left[ (j+1)q -1 \right]$ can be realised as descendants at level $rs$ of the vertex operator $e^{-i m \sqrt{2pq} \phi}$. As $m=-j,\cdots, j$ we have the $2j+1$ primary operators given by:
\bea 
\label{eqn:Construction}
\Phi_{(j)}^{m}= \CN(m) \CL_{rs} e^{-i \sqrt{2 p q} m \phi} 
\eea
where $ \CL_{rs}$ denotes the null vector at level $rs$ and $(r,s)=\left( (j+1-m)q-1 , (j+1+m)p-1 \right)$. We shall call this extra multiplet structure $\widetilde{SU(2)}$. It is an extra global symmetry of the chiral fields beyond the `minimal' sector.

We have also included a normalisation ${\mathcal{N}}(m)$. In cases in which the descendent field would vanish, for example $m=0$, we should take ${\mathcal{N}}(m)$ in a divergent way so that in the limit we get a non-trivial result. As we shall see this is equivalent to using the logarithmic null vectors \cite{Flohr:1998wm} rather than the normal ones.

We shall denote the vertex operator by:
\bea
V_m=e^{-im \sqrt{2pq}\phi}
\eea
which has conformal weight $h=m^2pq+m(p-q)$. We first note that:
\bea
\f{\p h}{\p m} = 2mpq + p-q
\eea
%
%This only vanishes in the case when $p=q=1$ (to see this note that $p$ and $q$ must have a common factor but as $gcd(p,q)=1$ this is when $p=q=1$) and also $m=0$.
The field $\f{\p V_m}{\p m}$ will not however any longer be a primary field as:
\bea
T(z) V_m (w) &\sim& \f{h V_m}{(z-w)^2} + \f{\p_w V_m}{z-w} \nonumber \\
T(z) \f{\p V_m}{\p m}(w) &\sim& \f{\f{\p h}{\p m} V_m+ h \f{\p V_m}{\p m}}{(z-w)^2} + \f{\p_w V_m}{z-w}
\eea
It will therefore be a logarithmic partner of $V_m$ (up to further subtleties if $\f{\p h}{\p m}=0$). Therefore in order to obtain the correct non-trivial primary field in these cases we should use the logarithmic null vectors instead of the normal ones in (\ref{eqn:Construction}). It is important to stress that although we are building our fields as descendants of the non-chiral operator $\f{\p V_m}{\p m}$ the final field we obtain \emph{will} be chiral. 

We also note that our choice of $\alpha=-m \sqrt{p~q}$ for the vertex operators is different from the one that is normally made in the Coulomb gas description described in the previous section. If we calculate $\alpha_{r,s}$ for the solutions (\ref{eqn:solns}) we obtain:
\bea
\alpha_{r,s}=\alpha_0+m \sqrt{pq}
\eea
which is not the same as the value $\alpha=-m \sqrt{p~q}$ (but of course gives rise to the same conformal dimension). This choice of the root of the quadratic equation for $\alpha_{r,s}$ is necessary in order to obtain the extra $SU(2)$ multiplet structure of the chiral fields.

Before illustrating this description of the chiral algebras we note that the $j=0$ case gives the Virasoro vacuum null vector at weight $h=(p-1)(q-1)$. Imposing the vanishing of this in all correlation functions gives us the standard `minimal' models \cite{Feigin:1992wv}. In order to describe rational extensions beyond this sector it is natural to conjecture that one must use these extra chiral algebras.

\subsection{Examples}
\subsubsection{$c_{1,1}=1$}
The first example that we shall analyse is the $c_{1,1}=1$ theory. In this case the stress tensor is simply:
\bea 
T=-\half \p \phi \p \phi
\eea
We have already given an example of the correlation functions of the triplet operators in the theory and that they should be identified with the $SU(2)_1$ generators. We shall now see that these are also reproduced by our free field construction.

In this case, and in all the $c_{p,1}$ models, the vacuum null vector vanishes corresponding to the fact that there is no `minimal' sector at all. Proceeding to first non-trivial case of the doublet fields, $j=\half$, we find:
\bea
\Phi_{(1/2)}^{-1/2}&=&e^{ \f{+i}{\sqrt{2}} \phi} \\
\Phi_{(1/2)}^{1/2}&=&e^{ \f{-i}{\sqrt{2}} \phi} \nonumber 
\eea
In this case $rs=0$ and we do not require any descendent structure. These fields have $h=\f{1}{4}$ and are parafermionic \cite{Fateev:1985mm}.

Now let us discuss the triplet, $j=1$, fields. The $\Phi_{(1)}^{\pm 1}$ cases are trivial as there is again no descendent structure. However for $m=0$ if $\CN(m)$ is regular then we will obtain $L_{-1} e^{0}=0$. Therefore, as we have already commented, the correct procedure to obtain a non-trivial result is to take $\CN(m)$ in a singular way. It is clear that choosing $\CN(m)=\f{1}{m}$ gives:
\bea
\Phi_{(1)}^{0}=\lim_{m \rightarrow 0} \CN(m) L_{-1} e^{-im \sqrt{2} \phi} =-i \sqrt{2} ~\p \phi 
\eea
This is certainly still a chiral field. Therefore the full set of currents at $j=1$ is given by:
\bea \label{eqn:c11W}
\Phi_{(1)}^{1}&=&e^{ +i \sqrt{2} \phi} \nonumber \\
\Phi_{(1)}^{0}&=& -i \sqrt{2} \p \phi \\
\Phi_{(1)}^{-1}&=&e^{ -i \sqrt{2} \phi} \nonumber 
\eea
These are the well known expressions for the $SU(2)_1$ generators. We have agreement with the results found earlier from the conformal blocks. 
\subsubsection{$c_{1,2}=-2$}
The second example that we shall discuss is the well studied case of the $c_{1,2}=-2$ model. In this case the stress tensor is given by:
\bea 
\label{eqn:cminus2stressT}
T=-\half \p \phi \p \phi + i \sqrt{2} \left( \f{-1}{2 \sqrt{2}} \right) \p^2 \phi=  -\half \p \phi \p \phi - \f{i}{2}  \p^2 \phi 
\eea
Again the first non-trivial chiral algebra, $j=\half$, is the doublet one at $h=1$:
\bea \label{eqn:cminus2doublets}
\Phi_{(1/2)}^{-1/2}&=&e^{ +i \phi} \\
\Phi_{(1/2)}^{1/2}&=& L_{-1} e^{ -i \phi} =  -i \p \phi e^{ -i \phi} \nonumber 
\eea
This is also a well known algebra - the symplectic fermions with OPEs:
\bea
\Phi_{(1/2)}^{1/2}(z) \Phi_{(1/2)}^{-1/2}(w) \sim \f{-1}{(z-w)^2}
\eea
In this case one can also write the stress tensor in terms of these operators:
\bea
T= \left(\Phi_{(1/2)}^{1/2} \Phi_{(1/2)}^{-1/2} \right)
\eea
However in general such an explicit display of these extra quantum numbers in the stress tensor does not seem possible.

In the next case of the triplet, $j=1$, fields the $\Phi_{(1)}^{\pm 1}$ components are simple:
\bea
\label{eqn:cminus2pm}
\phi_{(1)}^{-1}&=& e^{2i \phi} \nonumber \\
\phi_{(1)}^{1}&=& \left( L_{-2}+  \half L_{-1}^2 \right) e^{-2i \phi}
\eea
where $\left( L_{-2}-  \half L_{-1}^2 \right)$ is the level $2$ null vector of an $h=1$ state.

The $\Phi_{(1)}^{0}$ field requires the use of the level $3$ logarithmic null vector at $h=0$. The logarithmic pair is given by: $C=\f{-i}{2}$ and $D=\phi$. The factor of $\f{-i}{2}$ is necessary to ensure that they have the standard normalisation with respect to the stress tensor (\ref{eqn:cminus2stressT}):
\bea
T(z) C(w) &\sim& \f{\p C(w)}{z-w} \nonumber\\
T(z) D(w) &\sim& \f{C(w)}{(z-w)^2} + \f{\p D(w)}{z-w}
\eea
Now we can use the formula given in \cite{Flohr:1998wm} for the logarithmic null vector:
\bea
\label{eqn:cminus2w0}
\phi_{(1)}^{0}&=& \left( -L_{-1}^3+2 L_{-2} L_{-1} \right) D - L_{-3} C \nonumber\\
&=& - \p^3 \phi + 2 (T \p \phi) - \left( \f{-i}{2} \right) \p T \nonumber \\
&=& - \p \phi \p \phi \p \phi - \f{3}{2} i \p^2 \phi \p \phi + \f{1}{4} \p^3 \phi
\eea
where we use the convention of right normal ordering: $ABC=(A(BC))$. This is indeed chiral as claimed and it can be easily verified that this a primary operator of conformal weight $h=3$. Therefore in (\ref {eqn:cminus2pm},\ref{eqn:cminus2w0}) we have realised the full multiplet of $j=1$ fields.
\subsubsection{$c_{2,3}=0$}
This is one of the most important cases as it relates to the physically interesting situation of percolation \cite{Saleur:1992hk,Cardy:1992cm}. Many of the quantities of experimental and theoretical interest, such as crossing probabilities, are related to operators in the Kac-table. As the `minimal' model consists of only the vacuum $h_{1,1}=0$ an extended model is inevitable. 

The stress tensor is given by:
\bea
T=-\half \p \phi \p \phi + \f{i}{2 \sqrt{3}} \p^2 \phi
\eea
In this case we find the $j=\half$ doublet fields, at conformal weight $h=7$, are given by:
\bea \label{eqnczerofermdoublets}
\Phi_{(1/2)}^{1/2}&=& \CL_{6} e^{-i \sqrt{3} \phi}\\
\Phi_{(1/2)}^{-1/2}&=& \CL_{5} e^{i \sqrt{3} \phi} \nonumber 
\eea
The vertex operators in this case have $h=1$ and $h=2$ respectively. These generate a twisted $\CN=2$ algebra. Denoting $G^{\pm}=\sqrt{\f{2}{3}} e^{\mp i \sqrt{3} \phi}$ and $J=\f{-i}{\sqrt{3}} \p \phi$ we find:
\bea
G^+(z) G^-(w) &\sim& \f{2}{3 (z-w)^3} + \f{2 J(w)}{(z-w)^2} + \f{2 T(w)}{z-w} \nonumber \\
J(z) G^{\pm}(w) &\sim& \f{{\pm}G^{\pm}}{z-w}\\
J(z) J(w) &\sim& \f{1}{3(z-w)^2} \nonumber 
\eea
This was first suggested in \cite{Saleur:1992hk} to be the correct algebra to describe polymers and percolation theories containing certain non-trivial degenerate operators in the $c_{2,3}=0$ model. It is therefore very satisfying to see these same vertex operators come out of a more general description. In the $c=-2$ example the vertex operators $e^{\pm i \phi}$, used to build the symplectic fermions (\ref{eqn:cminus2doublets}), are \emph{not} themselves the chiral fields of the model. In a similar way we believe that the twisted $\CN=2$ fields are not the true chiral symmetries and that the full descendent structure in (\ref{eqnczerofermdoublets}) will be necessary. We shall discuss later the correct $W$-algebras in such models.
\section{Extended Chiral Algebras}
\subsection{Closure of the OPE}
We shall now show that the chiral fields $\Phi_{(j)}^m$ close amongst themselves as a $W$-algebra.

First we shall show that $h=\left(Np-1 \right) \left( Nq-1 \right)$ are the only dimensions that occur in both the $h_{r,1}$ and $h_{1,s}$ entries. We have:
\bea
h_{r,1}- h_{1,s} &=&0 \nonumber \\
(p-qs)^2-(pr-q)^2&=&0\\
(p-q+pr-qs)(p+q-pr-qs)&=&0 \nonumber 
\eea
Therefore we must have either $p(r+1)=q(s+1)$ or $p(1-r)=q(s-1)$. The second of these possibilities is ruled out by positivity of $r$ and $s$. Therefore we must have $p(r+1)=q(s+1)$. Now using $gcd(p,q)=1$ we conclude $r=Nq-1,s=Np-1$ with $N \in \mathbf{N}$. The case $N=1$ is within the `minimal' Kac-table and is not a generator of a chiral algebra. Therefore with the notation $N=2j+2$ we are reduced to the set that we have described before. Hence these are the only fields lying within both sets.

As our operators can lie within both the $h_{(2j+2)q-1,1}$ and $h_{1,(2j+2)p-1}$ it is now a simple exercise using the BPZ \cite{Belavin:1984vu} fusion rules\footnote{We of course are referring to the un-truncated ones as we are outside the minimal sector.} for the $h_{1,s}$ and $h_{r,1}$ operators separately to conclude that the operator algebra of these fields must close. As an example consider the $h=15$ fields from $c_{2,3}=0$. Regarding these as $h_{11,1}$ fields we have:
\bea 
\label{eqn:firstOPE}
h_{11,1} \otimes h_{11,1} = \left[h_{1,1}\right] + \left[h_{3,1}\right] + \cdots + \left[h_{21,1}\right]
\eea
The fields on the R.H.S. have dimensions: $\left\{ 0, \f{1}{3}, 2, 5, \f{28}{3}, 15, 22, \f{91}{3}, 40, 51, \f{190}{3} \right\}$. Where, unless stated otherwise, $\left[X\right]$ denotes contributions from $X$ and all its \emph{Virasoro} descendants.

Now regarding these fields as $h_{1,7}$ fields we have:
\bea \label{eqn:secondOPE}
h_{1,7} \otimes h_{1,7} = \left[h_{1,1}\right] + \left[h_{1,3}\right] + \cdots + \left[h_{1,13}\right]
\eea
where now the fields on the R.H.S. have dimensions: $\left\{ 0, 2, 7, 15, 26, 40, 57 \right\}$. As the OPE that we deduce must be the same from either (\ref{eqn:firstOPE}) or (\ref{eqn:secondOPE}) the only operators which can contribute are the ones present in both sequences which are in this case: $h=0, 2, 15, 40$.

By analysis of explicit forms of the solutions in this case \cite{Nichols:2002dk} we find that actually the identity operator at $h=0$ does \emph{not} appear in the OPE and the first contribution is from $h=2$. The chiral operator at $h=2$ is just the vacuum null vector (recall in general it is at $h=(p-1)(q-1)$). In all other examples we found that the first contributing operator was also the vacuum null vector rather than the identity. Using the fact that the vacuum null vector is just the $j=0$ case in our general construction we find the correct OPE for the chiral multiplets $\Phi_{(j)}$ is given by:
\bea
\Phi_{(j)} \otimes \Phi_{(j')}= \sum_{J=|j-j'|}^{j+j'} \left[ \Phi_{(J)} \right]
\eea
The existence of such extra $\widehat{SU(2)}$ structure in the fusion rules beyond the minimal sector should be expected from our construction (\ref{eqn:Construction}). In the case of the $c_{2,3}=0$ model this gives $h=2,15,40$ as described above. 

As these are chiral algebras then knowledge of the singular terms is sufficient. In the case of the $c_{2,3}$ example we cannot produce the $h=40$ operator in the singular terms of the OPE and thus we conclude the algebra is:
\bea
15 \otimes 15 =\left[ 1 \right] + \left[ 15 \right]
\eea
We have verified that this is true in this example by analysis of the possible rational solutions \cite{Nichols:2002dk,Nichols:2002qv}.

The triplet algebras, containing fields $W^a$, always close as $h_{(2)} > 2 h_{(1)}-1 $ and so lies beyond the singular terms. From the above analysis we can see that they must be of the form:
\bea
W^a \otimes W^b = g^{ab} \left[1 \right] + f^{ab}_c \left[W^c \right]
\eea
Again we have verified that this is true in all examples that we were able to check.

Although the algebra of the operators $\Phi_{(j)}$ will not close for $j>1$ we may take some subset of them, as in the orbifold constructions at $c=-2$ \cite{Kausch:1995py,Flohr:1996ea,Eholzer:1998se}, and get a closed algebra. We shall return to this point again later.
\subsection{Use of extra fermionic algebras}
 The use of additional chiral algebras to simplify fusion rules is well known in normal CFT \cite{Moore:1989ss}. We wish to show that within the $c=-2$ triplet model the fusion rules become significantly simpler when one takes into account the fermionic doublet algebra. However this can only be considered \emph{after} the rational model has been formed as otherwise the triplet fields are viewed as composites we do not have constraints from the associativity of the algebra.

The triplet algebra $W(2,3,3,3)$ in $c=-2$ is only associative if certain vacuum null vectors decouple \cite{Gaberdiel:1996np}. The vanishing of these null vectors in all correlators determines a consistent set of representations that close under fusion. The operator content is given by six fundamental representations: the irreducible ones : $\nu_0,\nu_1,\nu_{-1/8},\nu_{3/8}$ and the two indecomposable ones ${\mathcal R}_0,{\mathcal R}_1$. These have already been described many times in the literature (see for instance \cite{Gaberdiel:2001tr}). 

The full set of fusion rules have also been calculated \cite{Gaberdiel:1996np}:
\bea \label{eqn:oldfusionrules}
\nu_0 \otimes X &=&  X \quad \quad  {\rm for~ all ~ X }\nonumber \\
\nu_{-1/8} \otimes \nu_{-1/8} &=& \CR_0 \nonumber\\
\nu_{-1/8} \otimes \nu_{3/8} &=& \CR_1 \nonumber\\
\nu_{-1/8} \otimes \nu_1 &=& \nu_{3/8} \nonumber\\
\nu_{-1/8} \otimes \CR_0 &=& 2 \nu_{-1/8} \oplus 2 \nu_{3/8} \nonumber\\
\nu_{-1/8} \otimes \CR_1 &=& 2 \nu_{-1/8} \oplus 2 \nu_{3/8} \nonumber\\
\nu_{3/8} \otimes \nu_{3/8} &=& \CR_0 \nonumber\\
\nu_{3/8} \otimes \nu_1 &=& \nu_{-1/8}  \\
\nu_{3/8} \otimes \CR_0 &=& 2 \nu_{-1/8} \oplus 2 \nu_{3/8} \nonumber\\
\nu_{3/8} \otimes \CR_1 &=& 2 \nu_{-1/8} \oplus 2 \nu_{3/8} \nonumber\\
\nu_1 \otimes \nu_1 &=& \nu_0 \nonumber\\ 
\nu_1 \otimes \CR_0 &=& \CR_1 \nonumber\\
\nu_1 \otimes \CR_1 &=& \CR_0 \nonumber\\
\CR_0 \otimes \CR_0 &=& 2\CR_0 \oplus 2\CR_1  \nonumber\\
\CR_0 \otimes \CR_1 &=& 2\CR_0 \oplus \CR_1 \nonumber\\
\CR_1 \otimes \CR_1 &=& 2\CR_0 \oplus 2\CR_1\nonumber
\eea
We know from standard Virasoro fusion rules \cite{Belavin:1984vu} that:
\bea \label{eqn:BPZfuse}
h_{2,1} \otimes h_{2,1} = \left[ h_{1,1} \right] + \left[ h_{3,1} \right] 
\eea
Now, as we have already explained, to create a rational $c_{2,1}$ triplet model we extend the chiral algebra by the $h_{3,1}=3$ fields. Therefore with respect to the full chiral algebra all the terms on the R.H.S. of (\ref{eqn:BPZfuse}) are descendants of the unit operator $h_{1,1}$. In the $c=-2$ case the $h_{2,1}=1$ field is normally denoted $\nu_1$ and one can see in the fusion rules (\ref{eqn:oldfusionrules}) we do indeed have a closed sub-algebra:
\bea
\nu_1 \otimes \nu_1 &=& \nu_0
\eea
where $\nu_0$ is the identity representation. 

We can now consider splitting the total space into equivalence classes under the action of $\nu_1$ in the following way: if there exists a non-negative integer $n$ (actually $n=0,1$ is enough) such that:
\bea
\nu_1^n \otimes X=Y 
\eea
where $\nu_1^n$ denotes the $n^{th}$ fusion product $\nu_1 \otimes \cdots \otimes \nu_1$ then we put $X$ and $Y$ into the same equivalence class i.e.~  $X  \cong Y $. In the $c=-2$ case using the fusion rules (\ref{eqn:oldfusionrules}) we find:
\bea
\CR_0  &\cong& \CR_1  \nonumber\\
\nu_0  &\cong&  \nu_1  \\
\nu_{-1/8}  &\cong&  \nu_{3/8} \nonumber
\eea
We can also easily obtain the fusion rules for the equivalence classes:
\bea
 \nu_0  \otimes  X  &=&  X  \quad \quad  {\rm for~ all ~ X } \nonumber\\
 \nu_{-1/8}  \otimes  \nu_{-1/8}  &=&  \CR_0  \\
 \nu_{-1/8}  \otimes  \CR_0  &=& 4  \nu_{-1/8}  \nonumber\\
 \CR_0  \otimes  \CR_0  &=& 4  \CR_0 \nonumber
\eea
We stress that these are fusion rules for the chiral theory and not the local one. They are considerably simpler than those of the full triplet model and are essentially the fusion rules of the symplectic fermion model \cite{Kausch:2000fu}. In general the use of the extra closed fermionic algebra allows one to effectively reduce the number of basic fields present in the model. However they are not true chiral fields of the full model as they are not mutually local with respect to all operators and hence cannot be used to classify the model. We emphasize again that this can only be done \emph{after} the formation of a rational model so that the triplet fields are not viewed as composites.
\subsection{Vacuum character of triplet models}
We shall now demonstrate how, by a simple modification of the standard character formula in minimal models, we can obtain the Virasoro characters of our chiral fields. In this way we can produce the vacuum character of the theory and hence, by action of the modular group, all other characters.
 
The Virasoro character of fields $h_{r,s}$ in the minimal model is well known to be given by the Rocha-Caridi formula \cite{Rocha-Caridi}:
\bea
\chi^{Vir}_{rs}= \f{1}{\eta(q)} \sum_{n \in Z} \left[ q^{(2pqn+pr-qs)^2/4pq} - q^{(2pqn+pr+qs)^2/4pq} \right]
\eea
where we hope that no confusion will arise between the $q=e^{2 \pi i \tau}$ and the $q$ in the exponent. The Dedekind eta function is given by:
\bea
\eta(q) = q^{-1/24} \prod_{i=1}^{\infty} \left( 1-q^i \right)
\eea
Applying the Rocha-Caridi formula to our fields $h_{(2j+2)q-1,1}$ would give:
\bea \label{eqn:RC}
\chi^{Vir}_{(2j+2)q-1,1}&=& \f{1}{\eta(q)} \sum_{n \in Z} \left[ q^{(2pqn+2(j+1)pq-p-q)^2/4pq} - q^{(2pqn+2(j+1)pq-p+q)^2/4pq} \right] \nonumber \\
&&=\f{1}{\eta(q)} \sum_{m \in Z} \left[ q^{(2pqm-p-q)^2/4pq} - q^{(2pqm-p+q)^2/4pq} \right]
\eea
where we have made the substitution $m=n+j+1$. However we expect the correct character for the chiral fields at $h_{(j)}=\left[ (j+1)p-1 \right] \left[ (j+1)q -1 \right]$ to be of the form:
\bea
\f{1}{\eta(q)} \left[ q^{h_{(j)}} + \cdots \right]
\eea
where $\cdots$ denotes the contribution of \emph{higher} terms. The standard Rocha-Caridi formula (\ref{eqn:RC}) does not give this result as lower terms are also present. We therefore suggest a modification of this, removing these terms, for calculating the Virasoro character of the chiral fields $\Phi_j$ at weight $h_{(j)}$ in the extended $c_{p,q}$ model:
\bea \label{eqn:modifiedRC}
\chi^{Vir}_{(j)}&=& \f{1}{\eta(q)} \sum_{m \in Z} \left[ q^{(2pqm-p-q)^2/4pq} - q^{(2pqm-p+q)^2/4pq} \right] \nonumber \\
&& \quad - \f{1}{\eta(q)} \sum_{m=-j}^j \left[ q^{(2pqm-p-q)^2/4pq} - q^{(2pqm-p+q)^2/4pq}  \right]
\eea
Note that such an expression is invariant under exchanges of $p,q$. For the $c_{p,1}$ cases it reduces to the known expression with just one null vector \cite{Flohr:1996ea}.

Now the vacuum character of the triplet model is given by:
\bea 
\label{eqn:tripletc}
\chi^W_0= \sum_{j=0}^{\infty} (2j+1) \chi^{Vir}_{(j)}
\eea
where the $2j+1$ factor is due to the multiplicity of the $\Phi_{(j)}$ fields. The use of the standard Rocha-Caridi formula (\ref{eqn:RC}) for $\chi^{Vir}_{(j)}$ would make this expression divergent. Our modification in (\ref{eqn:modifiedRC}) removes these divergences and makes this $W$-vacuum character well behaved.

This triplet character (\ref{eqn:tripletc}) can be easily re-expressed in the form:
\bea
\chi^W_0= \sum_{m=-\infty}^{\infty} m^2 \left[ q^{(2pqm-p-q)^2/4pq} - q^{(2pqm-p+q)^2/4pq} \right]
\eea
Now introducing the forms:
\bea
\theta_{\lambda,k}&=& \sum_{n \in Z} q^{(2kn+\lambda)/4k} \\
\p \theta_{\lambda,k}&=&  \sum_{n \in Z} (2kn+\lambda) q^{(2kn+\lambda)/4k} \\
\p^2 \theta_{\lambda,k}&=&  \sum_{n \in Z} (2kn+\lambda)^2 q^{(2kn+\lambda)/4k} \\
\eea
we find the vacuum character can be expressed as:
\bea \label{eqn:tripletWchar}
\chi^W_0&=& \f{1}{4p^2q^2 \eta(q) } \left[ \p^2 \theta_{-p-q,pq}-2(-p-q) \p \theta_{-p-q,pq} + (-p-q)^2 \theta_{-p-q,pq} \right. \nonumber \\
&& \quad \left. - \p^2 \theta_{-p+q,pq}+2(-p+q) \p \theta_{-p+q,pq} - (-p+q)^2 \theta_{-p+q,pq}\right] 
\eea
We see, in contrast to the $c_{p,1}$ cases of \cite{Flohr:1996ea}, that this involves $\p^2 \theta$ and therefore will lead to $log^2(q)$ singularities (and therefore rank $3$ Jordan cells). From the general formula above one can see that the $c_{p,1}$ cases are special as the $\p^2 \theta$ terms cancel. At the point $c_{1,1}$ the $\p \theta$ terms also cancel and we are left with a non-logarithmic theory (namely $SU(2)_1$ WZNW).

Action of modular transformations on this character should lead to a closed set as in the case of $c_{p,1}$\footnote{We were kindly informed by M. Flohr of a basis of characters in the $c_{2,3}=0$ model closed under modular transformations. We have verified that our proposed vacuum character (\ref{eqn:tripletWchar}) can indeed be expressed as a linear combination of these.} but we shall not comment on this here \cite{WorkInProgress}.
\subsection{Orbifold models and their operator content}
Chiral fields used to classify a given model must be local with respect to all the operators of that model. We shall show from this simple consideration that the orbifold models naturally involve operators from the `fractional' entries in the Kac-table i.e. $h_{r,s}$ with $r,s$ not necessarily integer. We are not able to prove that such operators \emph{must} exist but it seems likely.

We first notice that the triplet fields based on the vertex operators $e^{\pm i \sqrt{2pq} \phi}$ are mutually local with respect to $V_{\pm}$ (\ref{eqn:CGscreenings}) and therefore with all operators based on the vertex operators $\alpha_{r,s}$ (\ref{eqn:CGalphas}) with $r,s \in N$. 

For the $Z_4$ model at $c=-2$ the correct algebra is $W(2,3,10^2)$ \cite{Eholzer:1998se}. The $h=10$ fields are based on the vertex operators $e^{\pm 2 i \sqrt{2pq} \phi}$. This is now local with respect to a larger set that is $\alpha_{r,s}$ with $r,s \in \half N$. It must be stressed that this is a formal notation for obtaining the conformal weights, from the use of (\ref{eqn:Kactableweights}), and is not related to the appearance of Virasoro degenerate operators. In a similar way the other $Z_n$ theories should involve operators with $2r,2s \in \mathbf{N}/n$ The appearance of operators with these conformal weights is in agreement with calculations of characters in the $\mathbf{Z}_n$ orbifold models of $c=-2$ \cite{Kausch:1995py,Flohr:1996ea,Kausch:2000fu}. It also suggests that in order to get a model at $c=0$ involving a $Z_4$ twisted sector \cite{Saleur:1992hk} we will require the algebra $W(2,15,40^2)$.

We should perhaps add a cautionary note that even at $c=-2$ the full operator content of any models beyond the triplet model is unknown. The characters of the $Z_n$ orbifold models suggest that the untwisted sector of this model is similar to that of the triplet model and in particular has only rank $2$ Jordan cells. However, as we explicitly demonstrate in the Appendix for the $h_{2,3}=0$ operators in the $c=-2$ model, it is simple to find correlation functions of operators in the Kac-table giving rise to higher rank Jordan cells. There seem to us to be several possible conclusions. Possibly there are no rational models that incorporate such operators, or the classification is incomplete, or perhaps the characters are misleading and do not give a correct indication of the fusion rules (as in recent examples of  $\widehat{SU(2)}$ at fractional level \cite{Gaberdiel:2001ny,Lesage:2002ch}). Clearly further work is required to clarify these issues. 
\section{Discussion}
As we have seen there is a rich class of chiral algebras in the $c_{p,q}$ models beyond the minimal sector. They naturally have an extra multiplet index transforming under an additional global $\widetilde{SU(2)}$ symmetry. Such structure in the the chiral sector will give rise to multiplet structure in many of the other fields of the model. It would be very interesting to find if there is some physical interpretation to these extra chiral symmetries. From the form of the vertex operators present in the chiral fields (\ref{eqn:Construction}) one would wish to interpret the extra quantum numbers as some kind of winding modes as is familiar in $c=1$. In the Coulomb gas however there is no obvious physical motivation for such a description. It seems possible that they correspond to some kind of quasi-particle excitations with $SU(2)$ quantum numbers. Such states are freely interacting as correlators can be calculated without the need for screening operators. In a heuristic way we have a strongly interacting gas of fundamental $h_{1,2}$ and $h_{2,1}$ `particles' and a certain point we have an abrupt change in the vacuum state from an empty one to one with quasi-particles. 

As the Coulomb gas in used is used to construct a wide range of conformal field theories similar extra structure should be expected. We have verified that this indeed happens with $SU(2)_{k}$ beyond the integrable representations. For $k+2=p/q$ we find exactly $2j+1$ rational solutions for the spin $J=(j+1)p-1$ fields. Moreover it can be shown that under hamiltonian reduction they reduce to the extended fields of the $c_{p,q}$ models - see \cite{Nichols:2001cv,Nichols:2002qv} for the example of $SU(2)_0$. Such results were expected as in \cite{Nichols:2002dk} the correlators in $c_{p,q}$ were produced in this way. It is an interesting question as to what happens when one considers representations of Lie groups larger than $SU(2)$ beyond the integrable representations. These naturally involve several Coulomb gas systems $(\phi_1,\cdots,\phi_n)$ and again it seems possible to produce extended chiral symmetries in a similar way but with larger multiplet symmetries. We hope to be able to give more extensive results soon.

The appearance of continuum chiral symmetries is suggestive that lattice models on which they are based may also possess some extra symmetry and it might prove productive to search more carefully for these. It also implies that the quantum groups must give rise to similar behaviour at the roots of unity \cite{Stanev:dr,Hadjiivanov:2000gx,Hadjiivanov:2001kr}.

In the minimal models it is of course possible, but extremely inefficient in practice, to work directly with the Virasoro vacuum null vector. There are many other, much simpler, ways to obtain the fusion rules and representation theory. In a similar sense it is clear that the explicit calculations of \cite{Gaberdiel:1996np} cannot realistically be extended and that it is necessary to develop the tools to understand the representation theory and embedding structure of these larger theories. Hopefully such work can help to clarify, and extend, many of the applications of these theories.
\section{Acknowledgements}
I am grateful to the EU for funding under the {\it Discrete Random Geometries: From Solid State Physics to Quantum Gravity} network. I would like to thank the IH\'ES, Paris for its warm hospitality where the final stages of this work were completed. I am grateful to H. Saleur, I. Kostov, D. Serban, I. Todorov and T. Popov for useful discussions.
\section{Appendix}
We consider here the $h_{2,3}=0$ operators in the $c_{2,1}=-2$ model. The null vector at level $6$ factorises in the following way:
\bea
\left. | \chi \right>= \left(L_{-3} -\f{2}{3} L_{-2}L_{-1} + \f{1}{12} L_{-1}^3 \right) \left(L_{-2} -\f{1}{2} L_{-1}^2 \right) L_{-1}\left. | h=0 \right>
\eea
Therefore the conformal blocks can be obtained by solving a third order equation and then two inhomogeneous equations of second and first order. The results for the conformal blocks of the correlator with four of these operators is:
\bea
F_1 &=& 1 \nonumber \\
F_2&=& \ln z \nonumber\\
F_3 &=& \ln(1-z) \\
F_4 &=& \ln^2 z  -2 \ln z \ln(1-z)\nonumber\\
F_5 &=& \ln^2(1-z) -2 \ln z  \ln(1-z) \nonumber\\
F_6 &=& \ln(1-z) \ln z + 2~ dilog(z) \nonumber
\eea
where the $dilog(z)$ function is given by:
\bea
dilog(z):= \int_1^x \f{\ln t}{t-1}
\eea
One can verify that in forming a single-valued correlator:
\bea
G(z,\bar{z}) = \sum_{i,j=1}^{6} U_{i,j} F_{i}(z) \overline{ F_j(z)} 
\eea
there are only $6$ possible solutions. Three of these correspond to correlators of operators in the triplet model (in the notation of \cite{Gaberdiel:1998ps} these are the primary operators $\mathbf{\Omega}$ and $\mathbf{\Theta^{\pm}}$ at $h=0$). However the other three are not as they imply a rank $3$ Jordan cell at $h=0$.

The simplicity of such solutions strongly suggests that it should be possible to obtain them by adding extra zero modes to the $c=-2$ system in a similar manner to \cite{Gurarie:1997dw,Krohn:2002gh}.

%\bibliographystyle{plain} % Alphabetical order
%\bibliographystyle{h-elsevier2} % Nuclear Physics format- no titles
%\bibliographystyle{unsrt}  % In order of reference with titles
%%
%\bibliography{adsrefs3}

\begin{thebibliography}{10}

\bibitem{Belavin:1984vu}
A.A. Belavin, A.M. Polyakov and A.B. Zamolodchikov,
\newblock Nucl. Phys. B241 (1984) 333.

\bibitem{Dotsenko:1984nm}
V.S. Dotsenko and V.A. Fateev,
\newblock Nucl. Phys. B240 (1984) 312.

\bibitem{Cardy:1992cm}
J.L. Cardy,
\newblock J. Phys. A25 (1992) L201, hep-th/9111026.

\bibitem{Gurarie:1993xq}
V. Gurarie,
\newblock Nucl. Phys. B410 (1993) 535, hep-th/9303160.

\bibitem{Rozansky:1993td}
L. Rozansky and H. Saleur,
\newblock Nucl. Phys. B389 (1993) 365, hep-th/9203069.

\bibitem{Rozansky:1992rx}
L. Rozansky and H. Saleur,
\newblock Nucl. Phys. B376 (1992) 461.

\bibitem{Bilal:1994nx}
A. Bilal and I.I. Kogan,
\newblock (1994), hep-th/9407151.

\bibitem{Caux:1997kq}
J.S. Caux et~al.,
\newblock Nucl. Phys. B489 (1997) 469, hep-th/9606138.

\bibitem{Kogan:1997nd}
I.I. Kogan, A. Lewis and O.A. Solovev,
\newblock Int. J. Mod. Phys. A12 (1997) 2425, hep-th/9607048.

\bibitem{Kogan:1997cm}
I.I. Kogan, A. Lewis and O.A. Solovev,
\newblock Nucl. Phys. Proc. Suppl. 56B (1997) 154, hep-th/9611208.

\bibitem{Giribet:2001qq}
G. Giribet,
\newblock Mod. Phys. Lett. A16 (2001) 821, hep-th/0105248.

\bibitem{Gaberdiel:2001ny}
M.R. Gaberdiel,
\newblock Nucl. Phys. B618 (2001) 407, hep-th/0105046.

\bibitem{Nichols:2001du}
A. Nichols,
\newblock Phys. Lett. B516 (2001) 439, hep-th/0102156.

\bibitem{Kogan:2001nj}
I.I. Kogan and A. Nichols,
\newblock Int. J. Mod. Phys. A17 (2002) 2615, hep-th/0107160.

\bibitem{Nichols:2001cv}
A. Nichols,
\newblock JHEP 0204 (2002) 056, hep-th/0112094.

\bibitem{Lesage:2002ch}
F. Lesage, P. Mathieu, J. Rasmussen and H. Saleur,
\newblock Nucl. Phys. B647 (2002) 363, hep-th/0207201.

\bibitem{Saleur:1992hk}
H. Saleur,
\newblock Nucl. Phys. B382 (1992) 486, hep-th/9111007.

\bibitem{Cardy}
J. Cardy,
\newblock (1999), cond-mat/9911024.

\bibitem{Gurarie:1999yx}
V. Gurarie and A.W.W. Ludwig,
\newblock J. Phys. A35 (2002) L377-L384, cond-mat/9911392.

\bibitem{Caux:1996nm}
J.S. Caux, I.I. Kogan and A.M. Tsvelik,
\newblock Nucl. Phys. B466 (1996) 444, hep-th/9511134.

\bibitem{Kogan:1996wk}
I.I. Kogan, C. Mudry and A.M. Tsvelik,
\newblock Phys. Rev. Lett. 77 (1996) 707, cond-mat/9602163.

\bibitem{Maassarani:1997jn}
Z. Maassarani and D. Serban,
\newblock Nucl. Phys. B489 (1997) 603, hep-th/9605062.

\bibitem{Gurarie:1997dw}
V. Gurarie, M. Flohr and C. Nayak,
\newblock Nucl. Phys. B498 (1997) 513, cond-mat/9701212.

\bibitem{CTT}
J. Caux, N. Taniguchi and A. Tsvelik,
\newblock Phys. Rev. Lett. 80 (1998) 1276, cond-mat/9711109.

\bibitem{Caux:1998eu}
J.S. Caux,
\newblock Phys. Rev. Lett. 81 (1998) 4196, cond-mat/9804133.

\bibitem{Bhaseen:1999nm}
M.J.Bhaseen, I.I.Kogan, O.A. Soloviev, N. Taniguchi and A.Tsvelik
,
\newblock Nucl. Phys. B580 (2000) 688, cond-mat/9912060.

\bibitem{Kogan:1999hz}
I.I. Kogan and A.M. Tsvelik,
\newblock Mod. Phys. Lett. A15 (2000) 931, hep-th/9912143.

\bibitem{Gurarie:1999bp}
V. Gurarie,
\newblock Nucl. Phys. B546 (1999) 765, cond-mat/9808063.

\bibitem{RezaRahimiTabar:2000qr}
M. Reza Rahimi~Tabar,
\newblock Nucl. Phys. B588 (2000) 630, cond-mat/0002309.

\bibitem{Bernard:2000vc}
D. Bernard and A. LeClair,
\newblock Phys. Rev. B64 (2001) 045306, cond-mat/0003075.

\bibitem{Bhaseen:2000bm}
M.J. Bhaseen,
\newblock Nucl. Phys. B604 (2001) 537, cond-mat/0011229.

\bibitem{Bhaseen:2000mi}
M. J. Bhaseen, J.-S. Caux, I. I. Kogan, and A. M. Tsvelik,
\newblock Nucl. Phys. B618 (2001) 465, cond-mat/0012240.

\bibitem{Ludwig:2000em}
A.W.W. Ludwig,
\newblock (2000), cond-mat/0012189.

\bibitem{CardyTalk}
J. Cardy,
\newblock (2001), cond-mat/0111031.

\bibitem{Kogan:2001ku}
I.I. Kogan and A. Nichols,
\newblock JHEP 01 (2002) 029, hep-th/0112008.

\bibitem{Kogan:1996df}
I.I. Kogan and N.E. Mavromatos,
\newblock Phys. Lett. B375 (1996) 111, hep-th/9512210.

\bibitem{Kogan:1996zv}
I.I. Kogan, N.E. Mavromatos and J.F. Wheater,
\newblock Phys. Lett. B387 (1996) 483, hep-th/9606102.

\bibitem{Ellis:1998bv}
J.R. Ellis, N.E. Mavromatos and D.V. Nanopoulos,
\newblock Int. J. Mod. Phys. A13 (1998) 5093, hep-th/9804084.

\bibitem{Ghezelbash:1998rj}
A.M. Ghezelbash, M. Khorrami and A. Aghamohammadi,
\newblock Int. J. Mod. Phys. A14 (1999) 2581, hep-th/9807034.

\bibitem{Kogan:1999bn}
I.I. Kogan,
\newblock Phys. Lett. B458 (1999) 66, hep-th/9903162.

\bibitem{Myung:1999nd}
Y.S. Myung and H.W. Lee,
\newblock JHEP 10 (1999) 009, hep-th/9904056.

\bibitem{Lewis:1999qv}
A. Lewis,
\newblock Phys. Lett. B480 (2000) 348, hep-th/9911163.

\bibitem{Nichols:2000mk}
A. Nichols and S. Sanjay,
\newblock Nucl. Phys. B597 (2001) 633, hep-th/0007007.

\bibitem{Kogan:2000nw}
I.I. Kogan and D. Polyakov,
\newblock Int. J. Mod. Phys. A16 (2001) 2559, hep-th/0012128.

\bibitem{Moghimi-Araghi:2001fg}
S. Moghimi-Araghi, S. Rouhani and M. Saadat,
\newblock Phys. Lett. B518 (2001) 157, hep-th/0105123.

\bibitem{Bakas:2002qh}
I. Bakas and K. Sfetsos,
\newblock Nucl.Phys. B639 (2002) 223-240, hep-th/0205006.

\bibitem{Jabbari-Faruji:2002xz}
S. Jabbari-Faruji and S. Rouhani,
\newblock Phys.Lett. B548 (2002) 237-242, hep-th/0205016.

\bibitem{RahimiTabar:1996dh}
M.R.R. Tabar and S. Rouhani,
\newblock Ann. Phys. 246 (1996) 446, hep-th/9503005.

\bibitem{RahimiTabar:1997nc}
M.R.R. Tabar and S. Rouhani,
\newblock Nuovo Cim. B112 (1997) 1079, hep-th/9507166.

\bibitem{Flohr:1996ik}
M.A.I. Flohr,
\newblock Nucl. Phys. B482 (1996) 567, hep-th/9606130.

\bibitem{RahimiTabar:1997ki}
M.R. Rahimi~Tabar and S. Rouhani,
\newblock Europhys. Lett. 37 (1997) 447, hep-th/9606143.

\bibitem{RahimiTabar:1996si}
M.R.R. Tabar and S. Rouhani,
\newblock (1996), hep-th/9606154.

\bibitem{Korchemsky:2001nx}
G.P. Korchemsky, J. Kotanski and A.N. Manashov,
\newblock Phys.Rev.Lett. 88 (2002) 122002, hep-ph/0111185.

\bibitem{Mahieu:2001iv}
S. Mahieu and P. Ruelle,
\newblock Phys.Rev. E64 (2001) 066130, hep-th/0107150.

\bibitem{Ruelle:2002jy}
P. Ruelle,
\newblock Phys.Lett. B539 (2002) 172-177, hep-th/0203105.

\bibitem{Cappelli:1997qf}
A. Cappelli, P. Valtancoli and L. Vergnano,
\newblock Nucl. Phys. B524 (1998) 469, hep-th/9710248.

\bibitem{Flohr:1998ew}
M.A.I. Flohr,
\newblock Phys. Lett. B444 (1998) 179, hep-th/9808169.

\bibitem{Rahimi-Tabar:1998ph}
M.R. Rahimi-Tabar and S. Rouhani,
\newblock Phys. Lett. B431 (1998) 85, hep-th/9707060.

\bibitem{Mavromatos:1998sa}
N.E. Mavromatos and R.J. Szabo,
\newblock Phys. Lett. B430 (1998) 94, hep-th/9803092.

\bibitem{Kogan:1998xm}
I.I. Kogan and A. Lewis,
\newblock Phys. Lett. B431 (1998) 77, hep-th/9802102.

\bibitem{Lewis:1998fg}
A. Lewis,
\newblock Nucl. Phys. B539 (1999) 367, hep-th/9808068.

\bibitem{Kogan:2000fa}
I.I. Kogan and J.F. Wheater,
\newblock Phys. Lett. B486 (2000) 353, hep-th/0003184.

\bibitem{Ishimoto:2001jv}
Y. Ishimoto,
\newblock Nucl. Phys. B619 (2001) 415, hep-th/0103064.

\bibitem{Kawai:2001ur}
S. Kawai and J.F. Wheater,
\newblock Phys. Lett. B508 (2001) 203, hep-th/0103197.

\bibitem{Bredthauer:2002ct}
A. Bredthauer and M. Flohr,
\newblock Nucl.Phys. B639 (2002) 450-470, hep-th/0204154.

\bibitem{RahimiTabar:1997ub}
M.R.R. Tabar, A. Aghamohammadi and M. Khorrami,
\newblock Nucl. Phys. B497 (1997) 555, hep-th/9610168.

\bibitem{Rohsiepe:1996qj}
F. Rohsiepe,
\newblock (1996), hep-th/9611160.

\bibitem{Kogan:1997fd}
I.I. Kogan and A. Lewis,
\newblock Nucl. Phys. B509 (1998) 687, hep-th/9705240.

\bibitem{Flohr:1998wm}
M.A.I. Flohr,
\newblock Nucl. Phys. B514 (1998) 523, hep-th/9707090.

\bibitem{Flohr:2000mc}
M. Flohr,
\newblock (2000), hep-th/0009137.

\bibitem{Flohr:2001tj}
M. Flohr,
\newblock Nucl.Phys. B634 (2002) 511-545, hep-th/0107242.

\bibitem{Tabar:2001et}
M.R.R. Tabar,
\newblock (2001), cond-mat/0111327.

\bibitem{Flohr:2001zs}
M. Flohr,
\newblock (2001), hep-th/0111228.

\bibitem{Gaberdiel:2001tr}
M.R. Gaberdiel,
\newblock (2001), hep-th/0111260.

\bibitem{Moghimi-Araghi:2002gk}
S. Moghimi-Araghi, S. Rouhani and M. Saadat,
\newblock (2002), hep-th/0201099.

\bibitem{Fjelstad:2002ei}
J. Fjelstad, J. Fuchs, S. Hwang, A.M. Semikhatov, and I.Yu. Tipunin,
\newblock Nucl.Phys. B633 (2002) 379-413, hep-th/0201091.

\bibitem{Mavromatos:2002fm}
N.E. Mavromatos and R.J. Szabo,
\newblock JHEP 01 (2003) 041, hep-th/0207273.

\bibitem{Krohn:2002gh}
M. Krohn and M. Flohr,
\newblock JHEP 01 (2003) 020, hep-th/0212016.

\bibitem{Kausch:1991vg}
H.G. Kausch,
\newblock Phys. Lett. B259 (1991) 448.

\bibitem{Kausch:1995py}
H.G. Kausch,
\newblock (1995), hep-th/9510149.

\bibitem{Flohr:1996ea}
M.A.I. Flohr,
\newblock Int. J. Mod. Phys. A11 (1996) 4147, hep-th/9509166.

\bibitem{Gaberdiel:1996np}
M.R. Gaberdiel and H.G. Kausch,
\newblock Phys. Lett. B386 (1996) 131, hep-th/9606050.

\bibitem{Flohr:1997vc}
M.A.I. Flohr,
\newblock Int. J. Mod. Phys. A12 (1997) 1943, hep-th/9605151.

\bibitem{Gaberdiel:1998ps}
M.R. Gaberdiel and H.G. Kausch,
\newblock Nucl. Phys. B538 (1999) 631, hep-th/9807091.

\bibitem{Kausch:2000fu}
H.G. Kausch,
\newblock Nucl. Phys. B583 (2000) 513, hep-th/0003029.

\bibitem{Nichols:2002dk}
A. Nichols,
\newblock JHEP 01 (2003) 022, hep-th/0205170.

\bibitem{Feigin:1992wv}
B.L. Feigin, T. Nakanishi and H. Ooguri,
\newblock Int. J. Mod. Phys. A7 (1992) 217.

\bibitem{Fateev:1985mm}
V.A. Fateev and A.B. Zamolodchikov,
\newblock Sov. Phys. JETP 62 (1985) 215.

\bibitem{Nichols:2002qv}
A. Nichols,
\newblock $SU(2)_k$ logarithmic conformal field theories, PhD thesis (2002), hep-th/0210070.

\bibitem{Eholzer:1998se}
W. Eholzer, L. Feher and A. Honecker,
\newblock Nucl. Phys. B518 (1998) 669, hep-th/9708160.

\bibitem{Moore:1989ss}
G.W. Moore and N. Seiberg,
\newblock Nucl. Phys. B313 (1989) 16.

\bibitem{Rocha-Caridi}
A. Rocha-Caridi,
\newblock \emph{Vacuum vector representations of the Virasoro algebra}, in \emph{Vertex Operators in Mathematics and Physics}, Springer-Verlag, 1985.

\bibitem{WorkInProgress}
A. Nichols, Work in Progress.

\bibitem{Stanev:dr}
Y. S. Stanev, I. T. Todorov and L. K. Hadjiivanov,
Phys. Lett. B276 (1992) 87.

\bibitem{Hadjiivanov:2000gx}
L. K. Hadjiivanov, Y. S. Stanev and I. T. Todorov,
Lett. Math. Phys. 54 (2000) 137, hep-th/0007187.

\bibitem{Hadjiivanov:2001kr}
L. Hadjiivanov and T. Popov,
\newblock hep-th/0109219.

\end{thebibliography}
% Include all (!!) the BIBTEX refs. \nocite{*}

\end{document}